\documentclass[twocolumn,showpacs,amsmath,amssymb]{revtex4}

\usepackage{graphicx}
\usepackage[latin1]{inputenc}
\usepackage{dcolumn}
\usepackage{bm}
\usepackage{epsfig}
\usepackage{color}

\begin{document}


\title{Visibility graphs and symbolic dynamics}

\author{Lucas Lacasa, Wolfram Just}
\affiliation{School of Mathematical Sciences, Queen Mary University of London, Mile End Road, E14NS London (UK)}%



\begin{abstract} 
Visibility algorithms are a family of geometric and ordering criteria by which a real-valued time series of $N$ data is mapped into a graph of $N$ nodes. This graph has been shown to often inherit in its topology nontrivial properties of the series structure, and can thus be seen as a combinatorial representation of a dynamical system. Here we explore in some detail the relation between visibility graphs and symbolic dynamics. To do that, we consider the degree sequence of {\it horizontal visibility graphs} generated by the one-parameter logistic map, for a range of values of the parameter for which the map shows chaotic behaviour. Numerically, we observe that in the chaotic region the block entropies of these sequences systematically converge to the Lyapunov exponent of the system. Via Pesin identity, this in turn suggests that these block entropies are converging to the Kolmogorov-Sinai entropy of the map, which ultimately suggests that the algorithm is implicitly and adaptively constructing phase space partitions which might have the generating property. To give analytical insight, we explore the relation $k(x), \ x\in[0,1]$ that, for a given datum with value $x$, assigns in graph space a node with degree $k$. 
In the case of the \textit{out}-degree sequence, such relation is indeed a piece-wise constant function. By making use of explicit methods and tools  from symbolic dynamics we are able to analytically show that the algorithm indeed performs an effective partition of the phase space and that such partition is naturally expressed as a countable union of subintervals, where the endpoints of each subinterval are related to the fixed point structure of the iterates of the map and the subinterval enumeration is associated with particular ordering structures that we called motifs.
\end{abstract}

\maketitle

\section{Introduction}
The family of visibility algorithms \cite{PNAS, PRE} are a set of simple criteria by which ordered real-valued sequences -and in particular, time series- can be mapped into graphs, thereby allowing the inspection of dynamical processes using the tools of graph theory. In recent years  research on this topic has essentially focused in two different fronts: from a theoretical perspective, some works have focused in providing a foundation to these transformations \cite{nonlinearity, Severini, theorem_bijection}, while in other cases authors have explored the resulting combinatorial analogues of some well-known dynamical measures \cite{EPL, PREToral}. Similarly, the graph-theoretical description of canonical routes to chaos \cite{PLOS, Route1, Route2, Route3} and some classical stochastic processes \cite{EPL, Ryan} have been discussed recently under this approach, as well as the exploration of relevant statistical properties such as time irreversibility \cite{EPJB,physio1}. From an applied perspective, these methods are routinely used to describe in combinatorial and topological terms experimental signals emerging in different fields including physics \cite{physics3,fluiddyn0,fluiddyn1,fluiddyn2,physics2,suyal,Zou}, neuroscience \cite{neuro,marinazzo,meditation_VG,motifs} or finance \cite{ryan1} to cite a few examples where analysis and classification of such signals is relevant.\\

\noindent Here we consider in some detail the so-called {\it horizontal visibility graph} (HVG) associated to paradigmatic examples of nonlinear, chaotic dynamics and we focus on a specific property of the graphs, namely the degree sequence, a set that lists the degree of each node. As will be shown below, HVGs inherit the time arrow of the associated time series and therefore their degree sequences are naturally ordered according to this time arrow. Since the degree of a node is an integer quantity, the degree series $\{k_t\}_{t=1}^{t_{max}}$ of an HVG can be seen as a integer representation of the associated time series $\{x_t\}_{t=1}^{t_{max}}$, i.e. as a symbolised series. However, such symbolization is far from trivial, as a priori there is no explicit partition of the state space which provides such a symbolization. 
In this work we explore this problem from the perspective of dynamical systems theory, and more particularly we  explore the connections between HVGs and symbolic dynamics. After briefly presenting the simple horizontal visibility algorithm in section II, in section III we explore the statistical properties of the degree sequence when constructed from a chaotic logistic map $x_{t+1}=rx_t(1-x_t)$.  We give numerical evidence that the block entropies over the degree sequence converge to the Lyapunov exponent of the map for all the values of the parameter $r$ for which the map shows chaotic behavior. Via Pesin theorem, this suggests that the Kolmogorov-Sinai entropy (a metric dynamical invariant of the map) finds a combinatorial analogue defined in terms of the statistics of the degree sequence.
This matching further suggests that the degree sequence is indeed produced after an effective symbolization of the system's trajectories. In section IV we explore by analytical means a possible effective partition of the phase space which could produce such symbolisation. Finally, in section V we close the article with some discussion.

\section{Horizontal visibility graphs}

\begin{figure}[h]
\centering
\includegraphics[width=0.95\columnwidth]{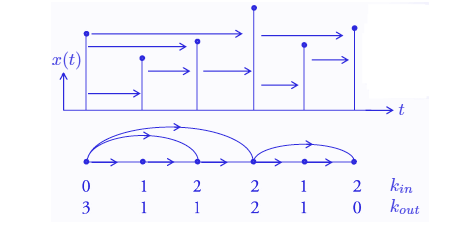}
\caption{Illustration of the process of constructing an 
horizontal visibility graph (HVG) from a time series. If one associates a time arrow to the links, we can decompose the degree of each node into an in-degree $k_{in}$ and an \textit{out}-degree $k_{out}$, making the HVG a directed graph.}
\label{HVG}
\end{figure}

The visibility algorithms \cite{PNAS, PRE} are a family of rules to map 
a real-valued time series $\{x_t\}_{t=1}^{t_\text{max}}, \ x_t \in \mathbb{R}$ 
into graphs (the multivariate version has been proposed recently \cite{Enzo}). 
In the \textit{horizontal visibility} case \cite{PRE}, each datum $x_t$ in 
the time series is associated with a vertex $t$ in the horizontal visibility 
graph (HVG) $\cal G$, and two vertices $i$ and $j$ are connected by a
directed edge in $\cal G$ if (see figure \ref{HVG})
\begin{equation}
x_k< \inf(x_i, x_j) \ \forall k:\  i<k<j.
\end{equation}
Geometrically, two vertices share an edge if the associated data are larger 
than any intermediate data. $\cal G$ can be characterized as a noncrossing 
outerplanar graph with a Hamiltonian path \cite{Severini}. Statistically, 
$\cal G$ is an order statistic \cite{Ryan} as it does not depend on the 
series marginal distribution.\\

\noindent Recent results suggest that the topological properties of $\cal G$ capture 
nontrivial structures of the time series, in such a way that graph theory 
can be used to describe classes of different dynamical systems in a succinct 
and novel way, including graph-theoretical descriptions of canonical routes 
to chaos \cite{Route1, Route2, Route3}, discrimination between stochastic 
and chaotic dynamics \cite{PREToral}. Interestingly, 
note that we can split the degree of a node 
$k_i=k^{\text{in}}_i + k^{\text{out}}_i$, where the \textit{in}-degree 
$k^{\text{in}}_i$ accounts for all the edges that node $i$ shares with nodes 
$j$ for which $j<i$ (so called past nodes), and where the \textit{out}-degree 
$k^{\text{out}}_i$ accounts for all the edges that node $i$ shares with nodes 
$j$ for which $j>i$ (so called future nodes). This splitting and its 
relation to the time arrow of the series allows to study the presence of 
time irreversibility in both deterministic and stochastic dynamical 
systems  \cite{EPJB,Ryan}.\\

\noindent Here, we pay particular attention to the degree sequences of $\cal G$ in 
the context of low dimensional chaotic dynamics, where 
$\{x_t\}_{t=1}^{t_\text{max}}, \ x_t \in [a,b]$. As commented before, the 
degree sequence of a graph is a set containing the degree $k$ of each node 
(total number of incident edges to a given node), $\{k_i\}_{i=1}^{t_\text{max}}$, 
where $k_i \in \mathbb{N}$, so it is a purely topological property of 
${\cal G}$. Empirical evidence suggests that this is a very informative 
feature and it was recently proved \cite{theorem_bijection} 
that, under mild conditions, there exists a bijection between the degree 
sequence and the adjacency matrix of an HVG. In other words the degree 
sequence often encapsulates all the complexity of the graph, so this 
sequence is a good candidate to account for the associated series complexity.\\

In HVGs, the time arrow indeed induces a natural ordering on the degree 
sequence where $k_i$ is the degree of node $i$ and nodes are ordered 
according to the natural time order. Thus, in some sense one may see 
$\{k_i\}$ as a coarse-grained symbolic representation of the time series. 
However, it is far from clear if $\{k_i\}$ results from any {\it effective} 
partitioning of the state space $[a,b]$ into a set of non-overlapping 
subsets: the algorithm itself does not partition the phase space 
explicitly. Furthermore, the number of different symbols $k_i$ is not 
determined a priori, as depending on the particular dynamics underlying the 
time series under study, the number of different degrees, i.e., the number 
of symbols might vary arbitrarily. Even worse, there does not seem to 
exist a unique transformation between the series datum $x_i$ and its 
associated node degree $k_i$: each $x_i$ may have a different associated 
symbol depending on the position of $x_i$ in the series. 
In this sense it is not straightforward at all to identify $\{k_i\}$ as a 
symbolic dynamics of the map. We shall explore these matters in detail, 
and we will show that we can indeed link a (non-standard) symbolic dynamics with the 
\textit{out}-degree sequence  $\{k^{\text{out}}_i\}_{i=1}^{t_\text{max}}$. Before we can explore these aspects in detail we want
to recall some background tools in symbolic dynamics, 
for the convenience of the reader. In addition, we will provide as well
some numerical evidence.

\section{Entropies}

\subsection{Symbolic dynamics and chaotic one dimensional maps}

The concept of entropy, originally introduced in thermodynamics, 
has become one of the most prominent measures to
quantify complexity. While there are numerous different notions
developed in the context of dynamical systems theory  
\cite{CE, CE2, Beck, Crutchfield}, the so-called
Kolmogorov-Sinai entropy is probably the prevalent concept.
Here we just summarise a couple of basic ideas which then will be used
as well in the graph theoretic setting.\\

\noindent \paragraph*{Partition and symbol sequence:} Consider a map on the interval $f: [a,b]\to [a,b]$, 
and denote by ${\cal P}=\{ {\cal I}_0,{\cal I}_1,\dots, {\cal I}_{n-1} \}$
a partition, i.e., a collection of pairwise disjoint sets
such that $$[a,b]= \bigcup_{j=0}^{n-1} {\cal I}_j.$$
The partition naturally induces a coding of $f$ \cite{CE2} defined by a 
map $\Phi: [a,b] \to \{0,1,\dots,n-1\}^{\mathbb{N}}$. In other words, 
it maps an initial value $x_0 \in [a,b]$ to an infinite symbol 
sequence $(\Phi_0(x_0),\Phi_1(x_0),\dots)$ , such that the $m$-th value in 
the symbol sequence $\Phi_m(x)=\sigma_m$ specifies the location of the $m$-th iterate of the map, $x_m=f^{(m)}(x_0)\in {\cal I}_{\sigma_m}$.\\ 

\paragraph*{Refinement, N-cylinders and generating partitions:} One can dynamically generate finer partitions
by using so-called $N$-cylinders \cite{Crutchfield, Beck}. Let us 
consider a finite symbol string 
$(\sigma_0,\sigma_1,\sigma_2,\dots,\sigma_{N-1} )$, where 
$\sigma_i$ can take any value from the alphabet of $n$ symbols. 
The set of initial values that generate this sequence, 
$J_{[\sigma_0,\sigma_1,\sigma_2,\dots,\sigma_{N-1}]}=
\{x_0 \in [a,b]: \Phi_i(x_0)=\sigma_i \ \forall i=0,1,\dots,N-1\}$ is called 
an $N$-cylinder of $f$. The ensemble of $N$-cylinders 
naturally induces another finer partition of $[a,b]$.
If in the limit $N\rightarrow \infty$
one finds that every $N$-cylinder contains at most a single 
point, then there is a correspondence between initial conditions $x \in [a,b]$ 
and symbol sequences. In that case, the original partition $\cal P$ 
is said to be generating.\\

\paragraph*{Block entropies:} To quantify the complexity of a dynamical system entropies are
the most prominent tools. If $\mu(\sigma_0,\sigma_1,\dots,\sigma_{N-1})$ 
denotes
the probability for the occurrence of a finite
symbol string $(\sigma_0,\sigma_1,\dots,\sigma_{N-1})$ then  
one defines the block entropy by 
\begin{eqnarray}
H_{\mu}({\cal P},N)&=&  -\sum_{\sigma_0,\sigma_1,\dots,\sigma_{N-1}} 
\mu(\sigma_0,\sigma_1,\dots,\sigma_{N-1})\cdot \nonumber \\ 
&&\cdot\ln[\mu(\sigma_0,\sigma_1,\dots,\sigma_{N-1})],
\label{metric}
\end{eqnarray}
where the summation is taken over all possible $N$-cylinders that can be 
formed according to a given partition ${\cal P}$. Under very mild
conditions Jensen's inequality guarantees the subadditivity 
of the quantity (\ref{metric}) and Fekete's lemma ensures the
existence of the limit
\begin{equation}
h_{\mu}=  \lim_{N\to \infty} \frac{H_{\mu}({\cal P},N)}{N}.
\label{hkss}
\end{equation}
A priori the value depends on the underlying partition
${\cal P}$. To remove this dependence one technically considers
the supremum over all possible partitions. If the partition is 
generating the value already coincides with the supremum and the
quantity is called the Kolmogorov-Sinai entropy.
Incidentally, note that it is more efficient to estimate numerically $h_\mu$ from 
$h_\mu= \lim_{N \to \infty}  [ H_{\mu}({\cal P},N) - H_{\mu}({\cal P},N-1) ]$
as, for a given partition, this latter formula converges faster with $N$ 
than eq.(\ref{hkss}). In addition, it is quite well established (see e.g. \cite{Pesin} or \cite{Ruelle}) that given
an absolutely continuous invariant measure 
the Kolmogorov-Sinai entropy and the (positive) 
Lyapunov exponent of the map coincide.\\

\begin{figure}[h]
\centering
\includegraphics[width=0.95\columnwidth]{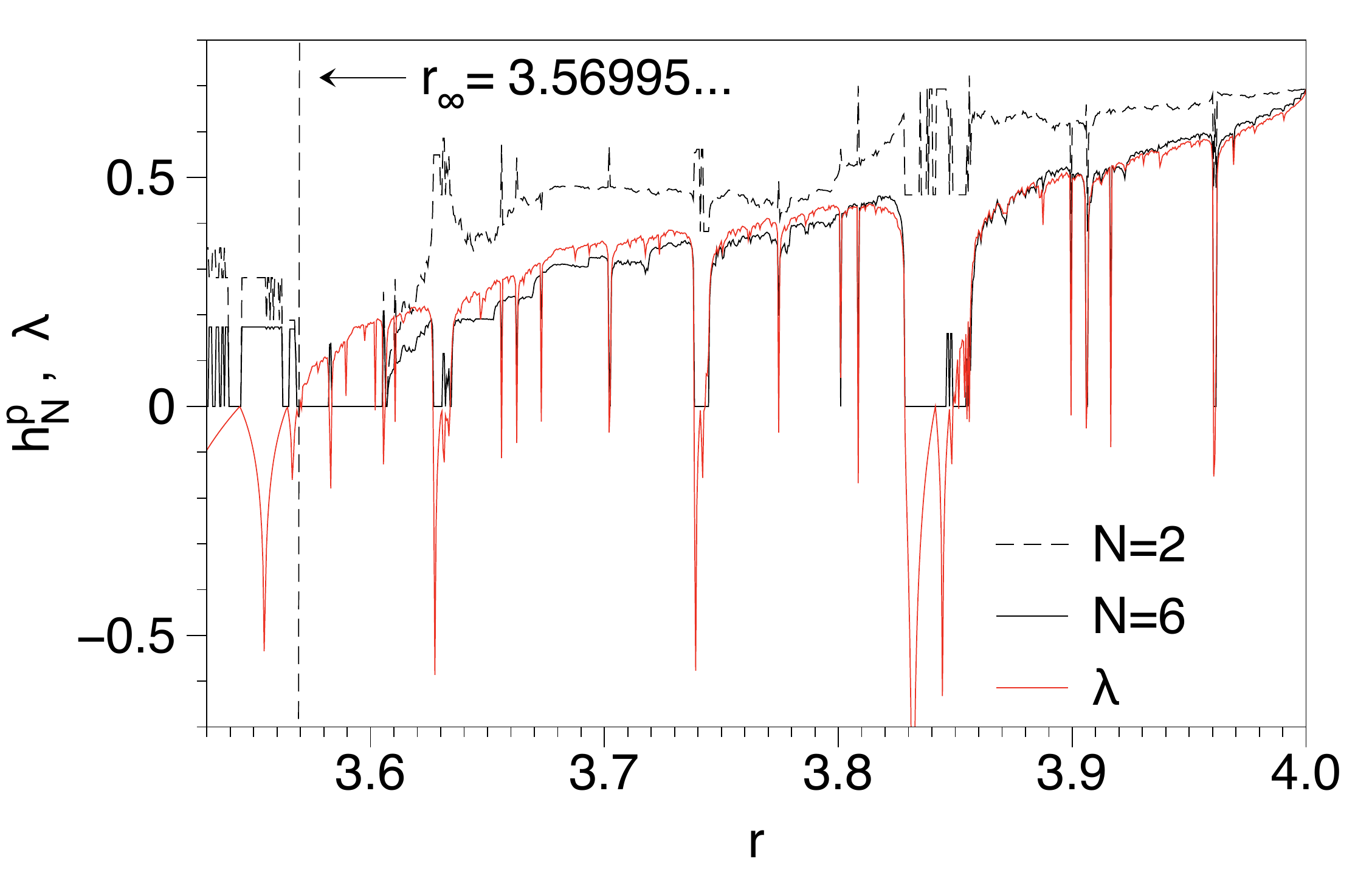}
\caption{Numerical estimate of the Kolmogorov-Sinai entropy and of the
Lyapunov exponent $\lambda$ for the logistic map $f(x)=rx(1-x)$ in 
dependence of
the bifurcation parameter $r$. Estimates of the entropy have been obtained
from $h^{\cal P}_N= H_{\mu}({\cal P},N) - H_{\mu}({\cal P},N-1)$ using different 
block sizes $N$ and a canonical partition with two symbols (see the text).}
\label{BE_S}
\end{figure}

\noindent From a practical point of view $H_{\mu}({\cal P},N)$ 
can be estimated numerically. 
One uses frequency histograms instead of the measure $\mu$, and takes into 
account that the estimations will only be accurate if $N$ is exponentially 
smaller than the series size \cite{Grassberger}. 
It is of capital importance to find out generating partitions. Unfortunately, 
there is no general strategy to 
determine whether a given partition is generating with the exception of 
axiom A systems \cite{Oono} and a few others \cite{Grassberger}. 
In this work we will focus on the logistic 
map \cite{Ledrappier} using
the canonical partition ${\cal P}=\{{\cal I}_L,{\cal I}_R\}$ with ${\cal I}_L=[0,1/2]$ and
${\cal I}_R=[1/2,1]$. With such a choice
numerical estimates of $h_\mu$ converge slowly from 
above \cite{Crutchfield}. 
For illustration, in figure \ref{BE_S} we have plotted the numerical 
results of $h^{\cal P}_{N}:=H_{\mu}({\cal P},N) - H_{\mu}({\cal P},N-1)$, estimated 
on symbolic sequences extracted from the logistic map $x_{t+1}=r x_t(1-x_t)$ with the canonical partition defined above.
The map displays both regular and chaotic dynamics as $r$ is varied. 
As the block size $N$ increases, $h_N$ approaches the Lyapunov exponent 
of the map $\lambda=\lim_{T\to \infty} \sum_{t=1}^T \ln|r-2r x_t|/T$,
for all values of $r$ for which $\lambda>0$, and $h_N \to 0$ otherwise. 
The approximation is quite bad for small 
values of $N$, whereas for large values of $N$, proper estimation 
of $\mu(\sigma_0\dots \sigma_{N-1})$ requires very long time series. 

\begin{figure*}[t]
\centering
\includegraphics[width=0.9\columnwidth]{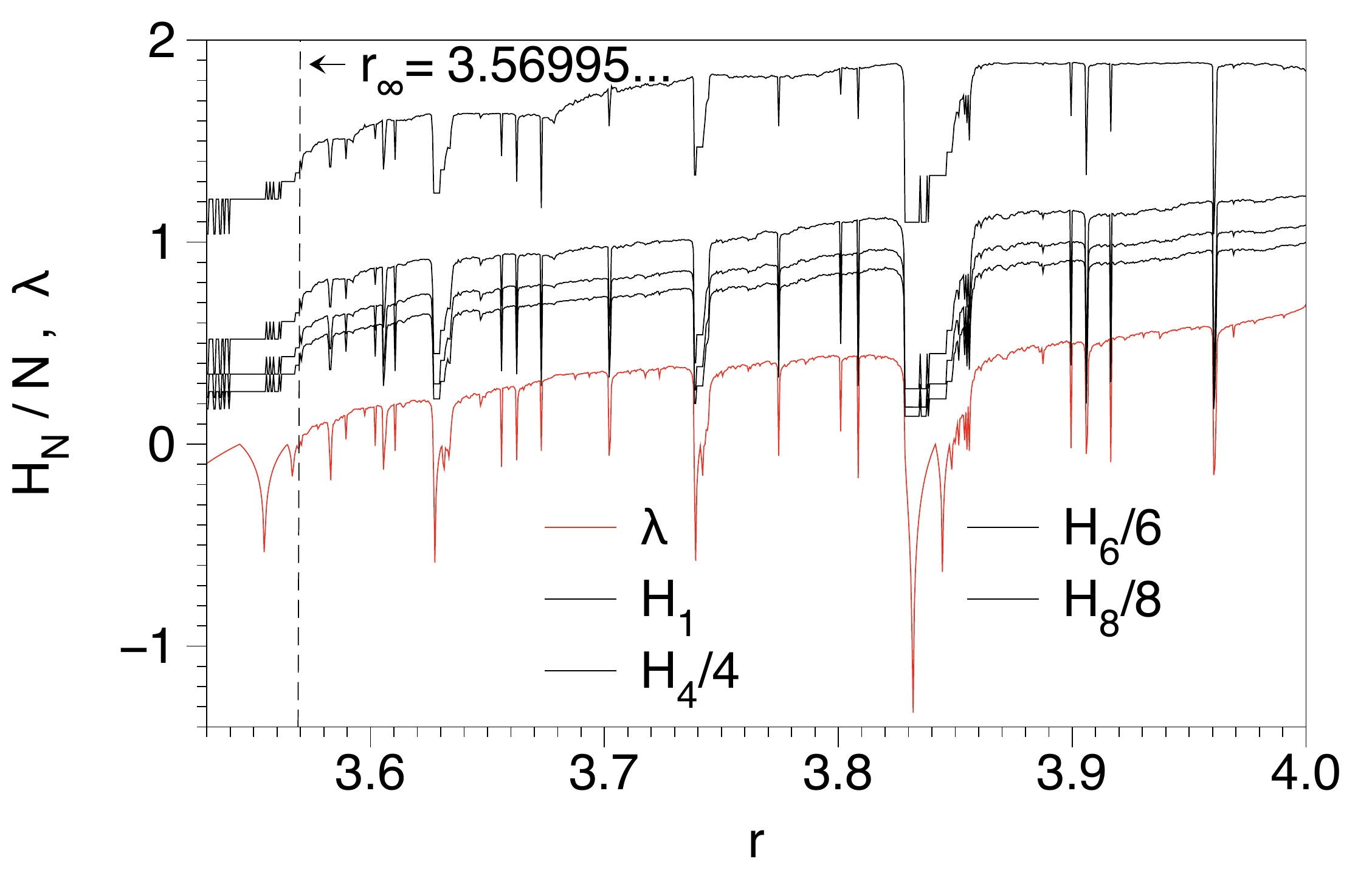}
\includegraphics[width=0.9\columnwidth]{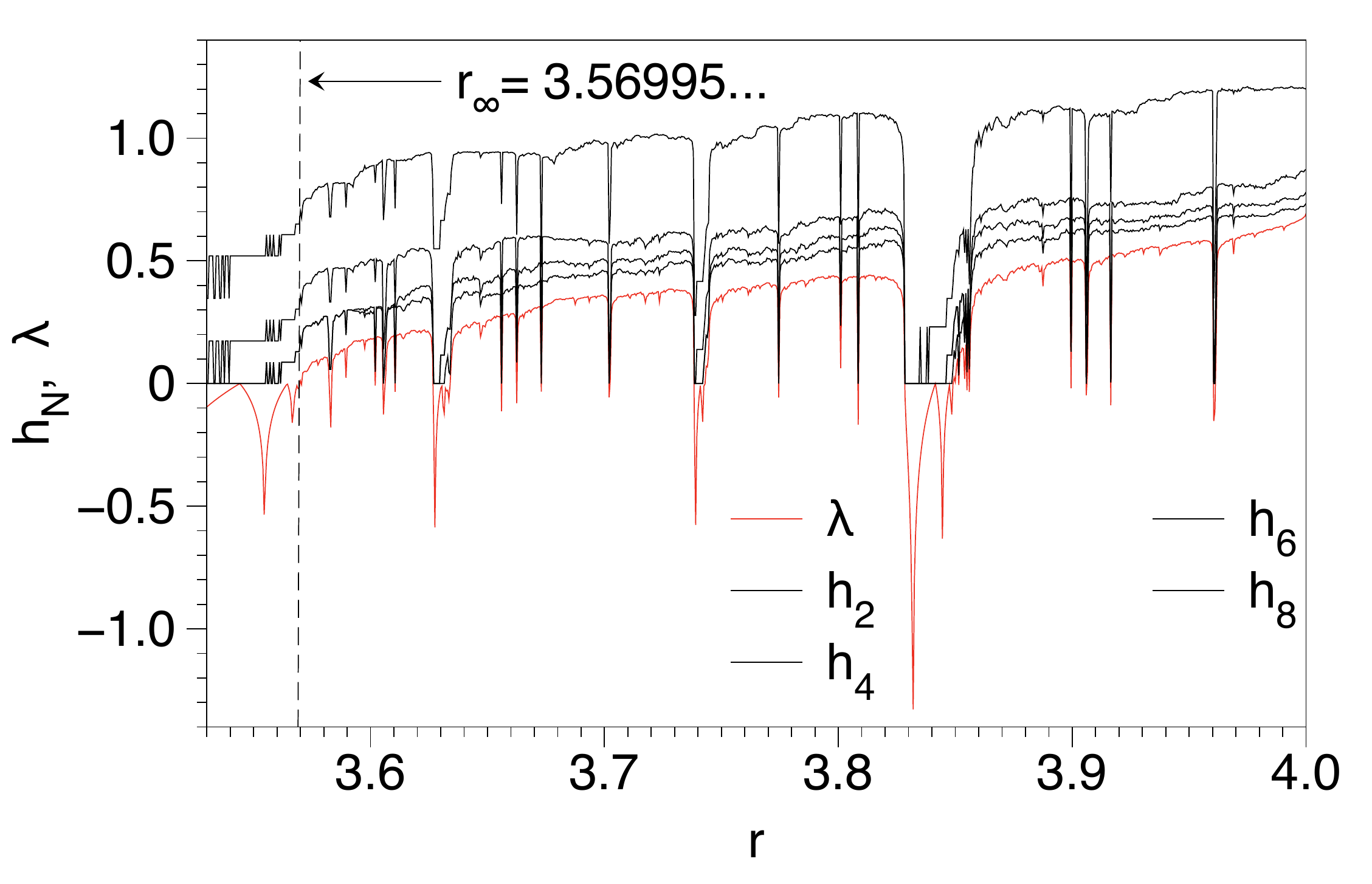}
\caption{Block entropies of the degree sequence $\{k_i\}$ 
as a function of the bifurcation parameter,
as obtained from the HVG of a time series of the logistic 
map $x_{t+1}=r x_t (1-x_t)$. Left: Normalised entropies $H_N/N$ 
for different block sizes $N$. Right:
Differential entropies $h_N=H_N-H_{N-1}$ for different block sizes $N$.
In addition, the figures contain
the Lyapunov exponent $\lambda$ of the map, cf.\ figure \ref{BE_S}.}
\label{Ent_K}
\end{figure*}

\subsection{Graph theoretic block entropies}

In direct analogy to the definition of a map's block entropy, see
eq.(\ref{metric}), we can define the HVG-block entropy as the block 
entropy of a particular degree sequence. In the case of the standard 
degree sequence $\{k_i\}$, this entropic quantity reads
\begin{equation}
H_N = -\sum_{\text{blocks}} P(k_1,\dots, k_N) \ln P(k_1,\dots, k_N),
\end{equation}
where the summation is performed over all admissible block strings 
of size $N$, $(k_1,\dots,k_N)$, and $P(k_1,k_2,\dots, k_N)$ denotes the
frequency of occurrence of the degree sequence $(k_1,\dots,k_N)$.
Similarly for the {\it out}-degree 
sequence $\{k_i^{\text{out}}\}$, we define
\begin{equation}\label{entout}
H^{\text{out}}_N = -\sum_{\text{blocks}} P(k^{\text{out}}_1,\dots, k^{\text{out}}_N) \ln P(k^{\text{out}}_1,\dots, k^{\text{out}}_N) \, .
\end{equation}
The differential block entropy $h_N$ is defined in direct equivalence to 
its counterpart in the map, $h_N=H_N - H_{N-1}$ and we are again interested in 
the limits $\lim_{\rightarrow \infty} H_N/N$ and $\lim_{N\rightarrow \infty} h_N$,
see figure \ref{Ent_K}. Fortunately, Jensen's inequality and Fekete's lemma
ensures the existence of limits, as before.\\

\noindent At a phenomenological level, the numerical results displayed in figure \ref{Ent_K} suggest that the entropy defined
on the basis of node degree sequences shows striking similarity to the
entropy defined with respect to generating partitions of the phase space.
In particular, we obtain expressions which seem to converge in
a monotonic decreasing way towards the Lyapunov exponent of the map (and this also holds for the {\it out}-degree sequence, see figure \ref{BEout}).
In fact, it has been pointed out recently \cite{Route1, PLOS} that $H_1$, 
which is nothing but the Shannon entropy over a graph's degree distribution 
$P(k)$, is qualitatively similar to the Lyapunov exponent $\lambda$  in the 
Feigenbaum scenario. Hence, it seems sensible to investigate to which 
extent the degree sequence shares properties of symbolic dynamics.
Among others we will discuss whether the degree sequence
provide a partition of the phase space, whether this partition finally 
determines a phase space point, and whether 
the statistical properties of the degree sequence give additional 
nontrivial insight into the dynamics of the map $f$.

\section{The horizontal visibility algorithm and symbolic dynamics}

For the following considerations we will focus exclusively
on the fully chaotic logistic map $f(x)=4 x(1-x)$. Large parts of
our considerations apply to general parameter values $r<4$
if a suitable pruning of the symbolic dynamics will be taken into
account. Our main concern is the question whether the degree sequence
induces an effective partition of phase space. For this purpose we
will first clarify
to which extent node degrees can be considered as functions of the
initial condition and whether the properties of such a function
are amenable to a theoretical investigation.

\subsection{Node degrees as phase space functions}

\begin{figure*}[t]
\centering
\includegraphics[width=0.9\columnwidth]{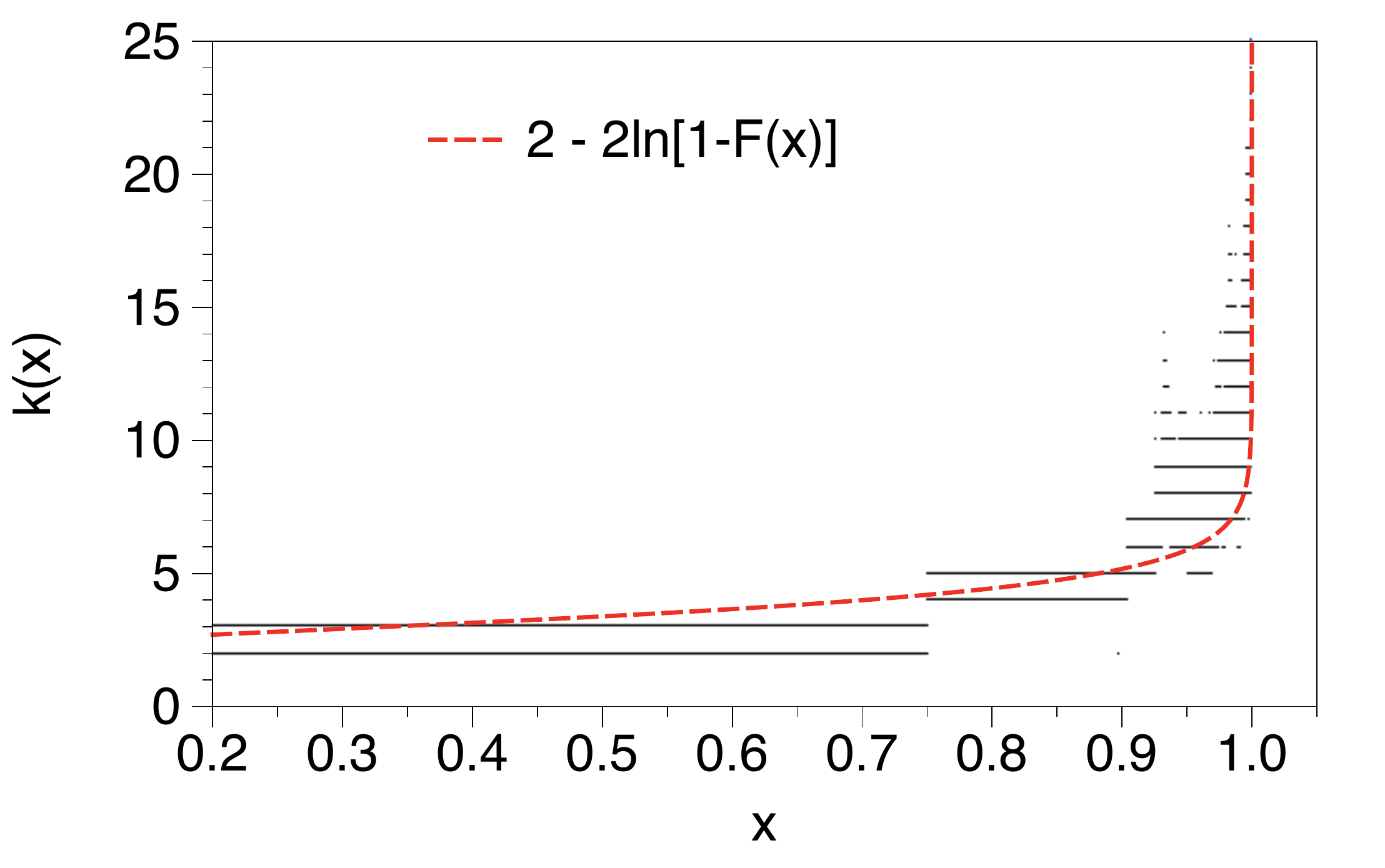}
\includegraphics[width=0.9\columnwidth]{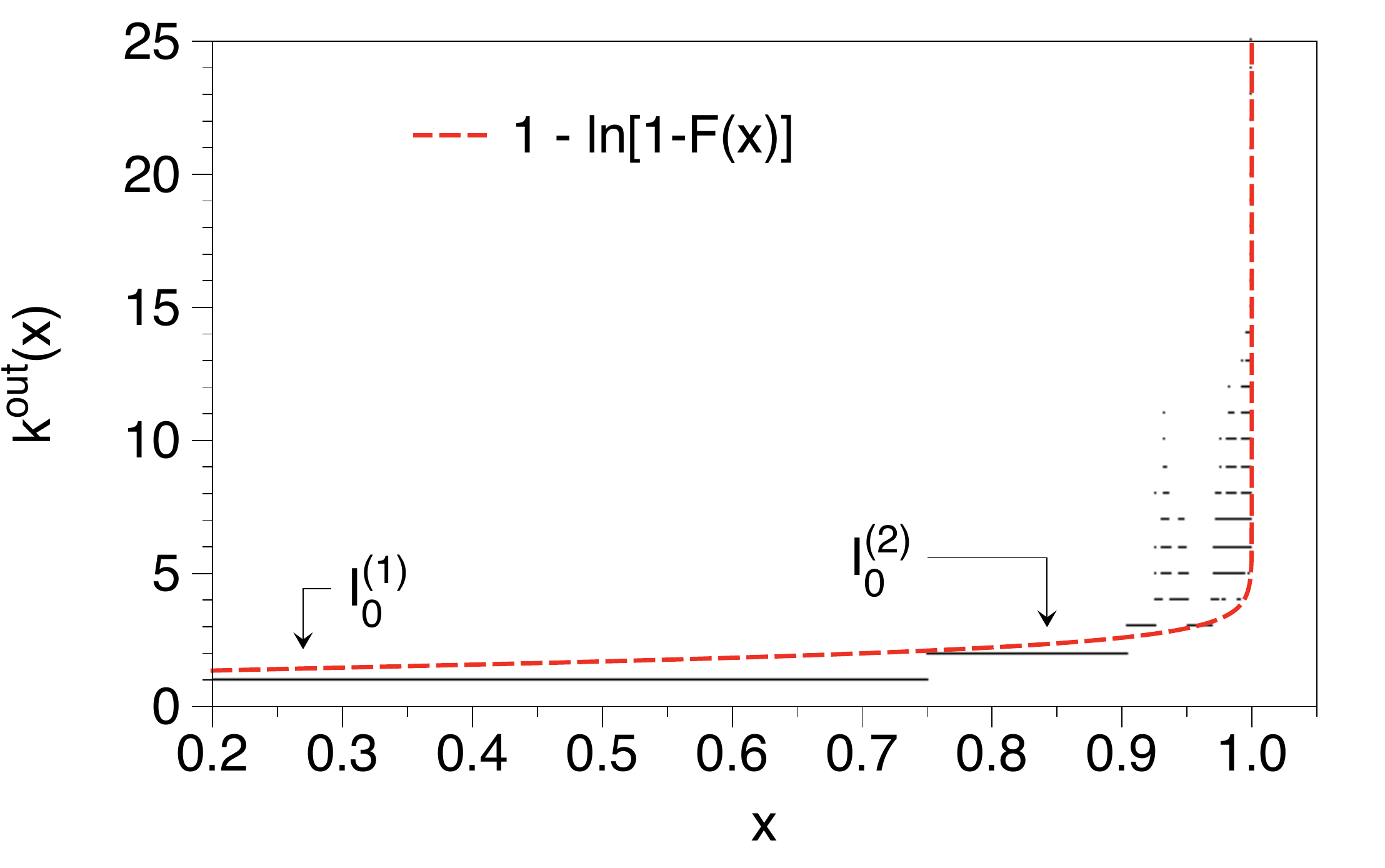}
\caption{Node degree $k(x)$ in the case of a fully chaotic ($r=4$) logistic 
map as obtained from the HVG of a time series (left). Analogous plot
for the  out-degree $k^{\text{out}}(x)$ (right).
The labels used for enumerating the piecewise constant branches will be 
introduced in the main text. Broken lines show the analytical estimates, eqs.(\ref{apa}) and (\ref{apb}).}
\label{part_k}
\end{figure*}

Given a time series, the node degree $k_t$ of a datum $x_t$ is easily obtained from the degree sequence of the graph.
We can therefore numerically reconstruct $k(x)$ which plots the node degree as a function of the phase space 
coordinate. We have generated both the HVG and its directed version, associated to a time series of $10^5$ data from a fully chaotic ($r=4$) logistic map. In the left panel of figure \ref{part_k} we then plot $k(x)$, whereas in the right panel of the same figure the corresponding \textit{out}-degree function $k^{\text{out}}(x)$ is plotted.

\begin{figure}[h]
\centering
\includegraphics[width=0.95\columnwidth]{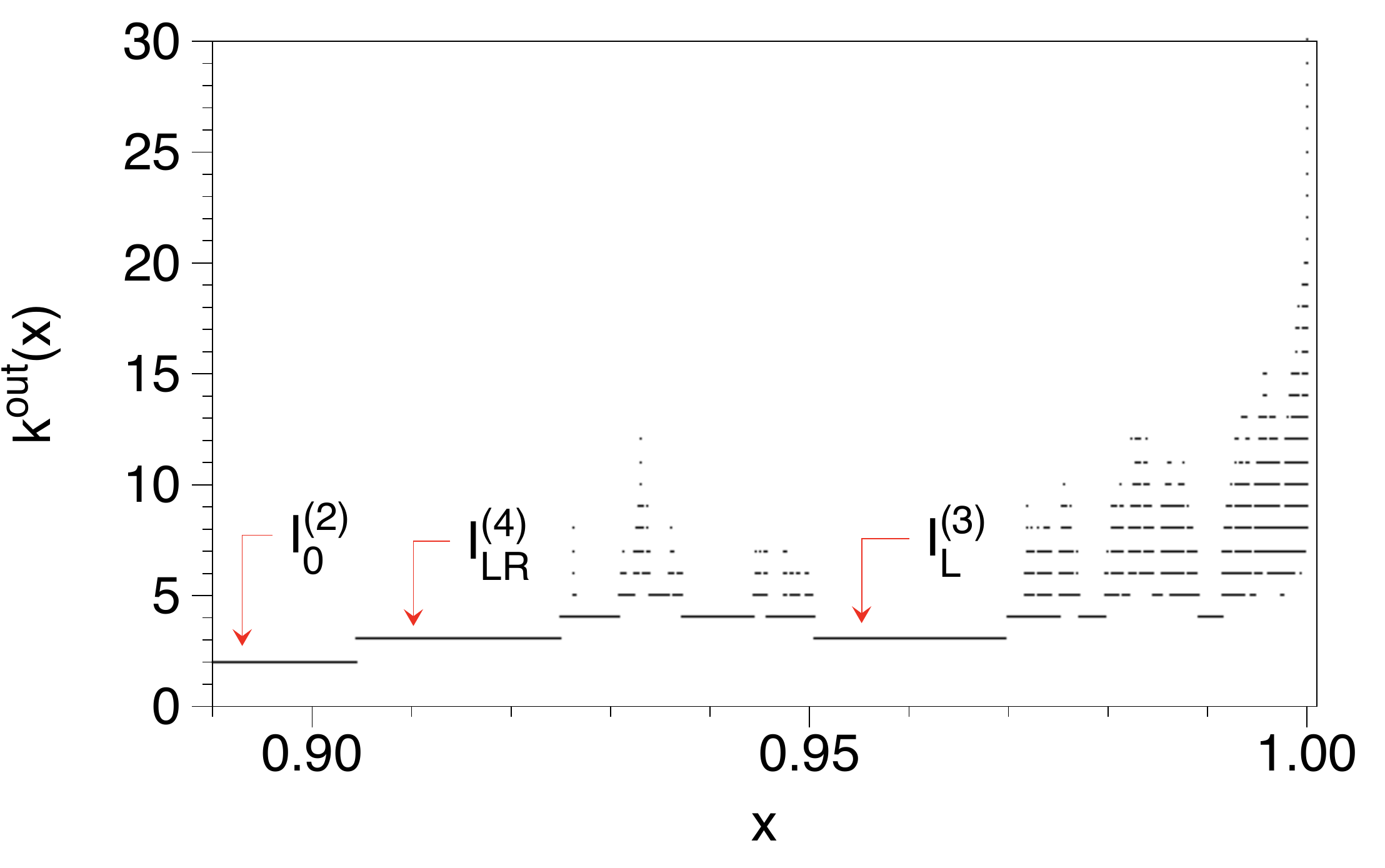}
\caption{A detailed view of $k^{\text{out}}(x)$ for the fully chaotic
logistic map (cf. figure \ref{part_k}). The labels used for enumerating 
the piecewise constant branches will be introduced in the main text. Vertical lines correspond to initial conditions which result in motifs with an unbounded \textit{out}-degree and and unbounded number of hidden nodes, see the main text.}
\label{zoom}
\end{figure}

\noindent As expected, $k(x)$ is a multivalued function since the logistic map
is not invertible. Interestingly, $k^{\text{out}}(x)$ seems to be
single-valued, although the shape is highly heterogeneous, particularly in the region closer to the upper bound of the interval. 
A zoom of this plot close to $x=1$ is depicted in figure \ref{zoom}, 
highlighting its complex structure. Before we proceed to evaluate this
intricate structure, let us first focus on the overall trend of
the two functions which can be captured by a simple stochastic argument.
Assuming that we can neglect correlations in the chaotic time series
we can model such a series as a set of uncorrelated random 
variables with probability distribution $\rho(x)=[\pi\sqrt{x(1-x)}]^{-1}$ (i.e. the invariant measure of the map). 
It has been proven \cite{PRE} that under such assumptions, the degree 
distribution conditioned to $x$ is given by
\begin{equation}
P(k|x)= \sum_{j=0}^{k-2}\frac{(-1)^{k-2}}{j!(k-2-j)!}
[1-F(x)]^2\{\ln[1-F(x)]\}^{k-2}.
\end{equation}
A similar expression applies for the \textit{out}-degree distribution \cite{nonlinearity}, 
\begin{equation}
P(k^{\text{out}}|x)=P(k^{\text{in}}|x)= \frac{(-1)^{k-1}}{(k-1)!}
[1-F(x)]\{\ln[1-F(x)]\}^{k-1},
\end{equation}
where $F(x)=\int_0^x \rho(x')dx'$ is the cumulative distribution
function of the underlying density. 
Accordingly, one can extract  an 'average function' which for 
the uncorrelated series reads
\begin{equation}\label{apa}
\langle k(x)\rangle = \sum_{k=2}^\infty k P(k|x)=2-2\ln(1-F(x))
\end{equation}
and
\begin{equation}\label{apb}
\langle k^{\text{out}}(x)\rangle = \langle 
k^{\text{in}}(x)\rangle =\sum_{k=1}^\infty k P(k^{\text{out}}|x)=1-\ln(1-F(x)).
\end{equation}
A comparison of eqs.(\ref{apa}) and (\ref{apb})
with the numerical 
estimates of $k(x)$ and $k^{\text{out}}(x)$ is shown in figure \ref{part_k} 
with $F(x)=1/2 +\arcsin(2x-1)/\pi$. 
It is interesting to see that 
the analytic prediction for the logistic map 
produces large values for
the node degree only in the vicinity of $x=1$.\\

\subsection{Motifs}
\begin{figure}[h]
\centering
\includegraphics[width=0.99\columnwidth]{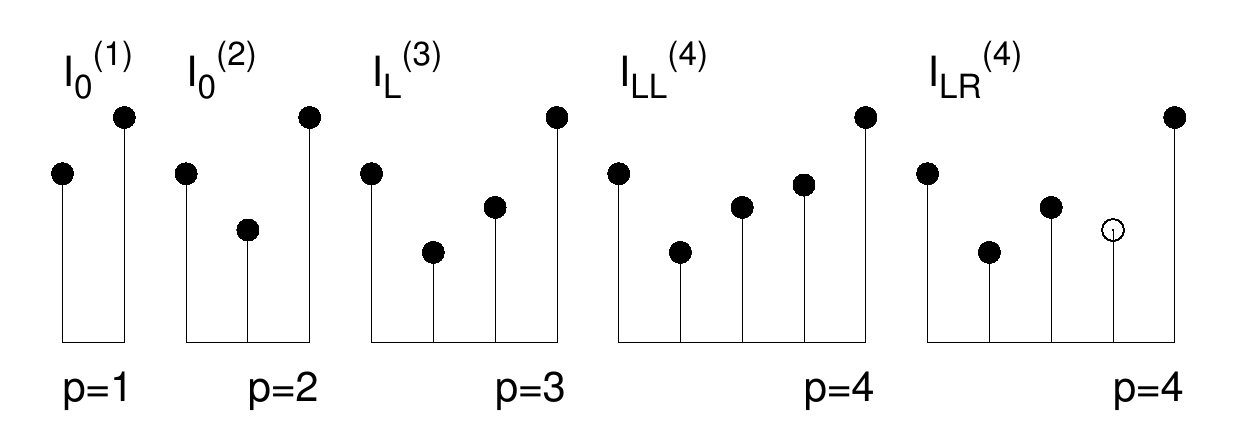}
\caption{A sketch of all motifs of the fully chaotic logistic map
up to length $p=4$. Visible/hidden nodes are indicated by full/open
symbols (cf. figure \ref{HVG}). The labels will be introduced and explained in the
main text.}
\label{motive}
\end{figure}

A closer look at $k^{\text{out}}(x)$ reveals that this is
a piecewise constant function of the initial condition,
see figure \ref{zoom}. In fact, such a property is not
really a surprise since any initial condition determines uniquely a 
forward orbit and thus all the outgoing links from the seed node. 
With a slight abuse of notation, we call the subgraph obtained from an orbit of $p+1$ data such that there is a link between the initial and final data a 
graph motif (note that $p$ considers both the visible and  hidden nodes as well, see figure \ref{motive} for illustration). The concept is closely related but not totally identical to the so-called sequential motifs introduced recently \cite{motifs}.\\
In formal terms, a motif is a (finite)
sequence of $p+1$ orbit points $(x_0,x_1,\ldots,x_p)$ such that $x_0\leq x_p$
and $x_0> x_\ell$ for $1\leq \ell \leq p-1$. For an orbit of $p+1$ data, we say that the length of the associated motive is $p$.
The ordering of the intermediate
points $x_\ell$ determines the visibility, the outgoing links, and thus
the \textit{out}-degree $k^{\text{out}}(x_0)$. By studying motifs we are therefore
able to uncover the structure of the function $k^{\text{out}}(x)$ as follows: we will first locate the sets of initial conditions which give rise to particular motif. These sets will be shown to be subintervals labelled ${\cal I}_{\bf x}^{(p )}$ for the motif of length $p$, where the specific labeling $\bf x$ will be made evident below (see figure \ref{motive} for all motifs of length $p\leq 4$). We will be able to associate a specific \textit{out}-degree to each subinterval ${\cal I}_{\bf x}^{(p )}$ such that
$$[0,1]=\bigcup_{p,{\bf x}} {\cal I}_{\bf x}^{(p )}.$$
This phase space partition will then yield, in principle, a way to reconstruct $k^{\text{out}}(x_0)$ and thus build up the effective symbolisation that we are looking for.

\subsection{motifs of short length $p\leq 4$}

We start by considering all admissible motifs up to length $p=4$, these are sketched in figure \ref{motive}. Note that some motifs are not appearing, for instance the one for which $x_0>x_1>x_2,\ x_0<x_3$ is forbidden as for the fully chaotic logistic map it is not possible to find three consecutive data in monotonically decreasing order. In what follows we will explicitly compute the sets of initial conditions that give rise to these motifs along with the relevant notation.

\subsubsection{$p=1$:} Given the logistic map $f(x)=4x(1-x)$ the set of points
which obeys $x_0<f(x_0)=x_1$ is simply the interval $${\cal I}_0^{(1)}=[0,\xi_2^{(1)}]$$
bounded by the nontrivial fixed point $\xi_2^{(1)}=3/4$ of the map.
Here we use the notation $\xi_k^{(p)}$ to denote the fixed points
of the $p$-th iterate, $f^{(p)}$, sorted by size, where $k$ runs
from $1$ to $2^p$. Hence $k^{\text{out}}(x)$ takes the value $1$ on the interval
${\cal I}_0^{(1)}$,  see figure \ref{part_k}. In fact, the intervals $[0,\xi_2^{(1)}]$ 
and $[\xi_2^{(1)},1]$  provide a (non generating) Markov partition and the latter interval 
is mapped onto the former by the map $f$.

\subsubsection{$p=2$:} 
\begin{figure}[h]
\centering
\includegraphics[width=0.99\columnwidth]{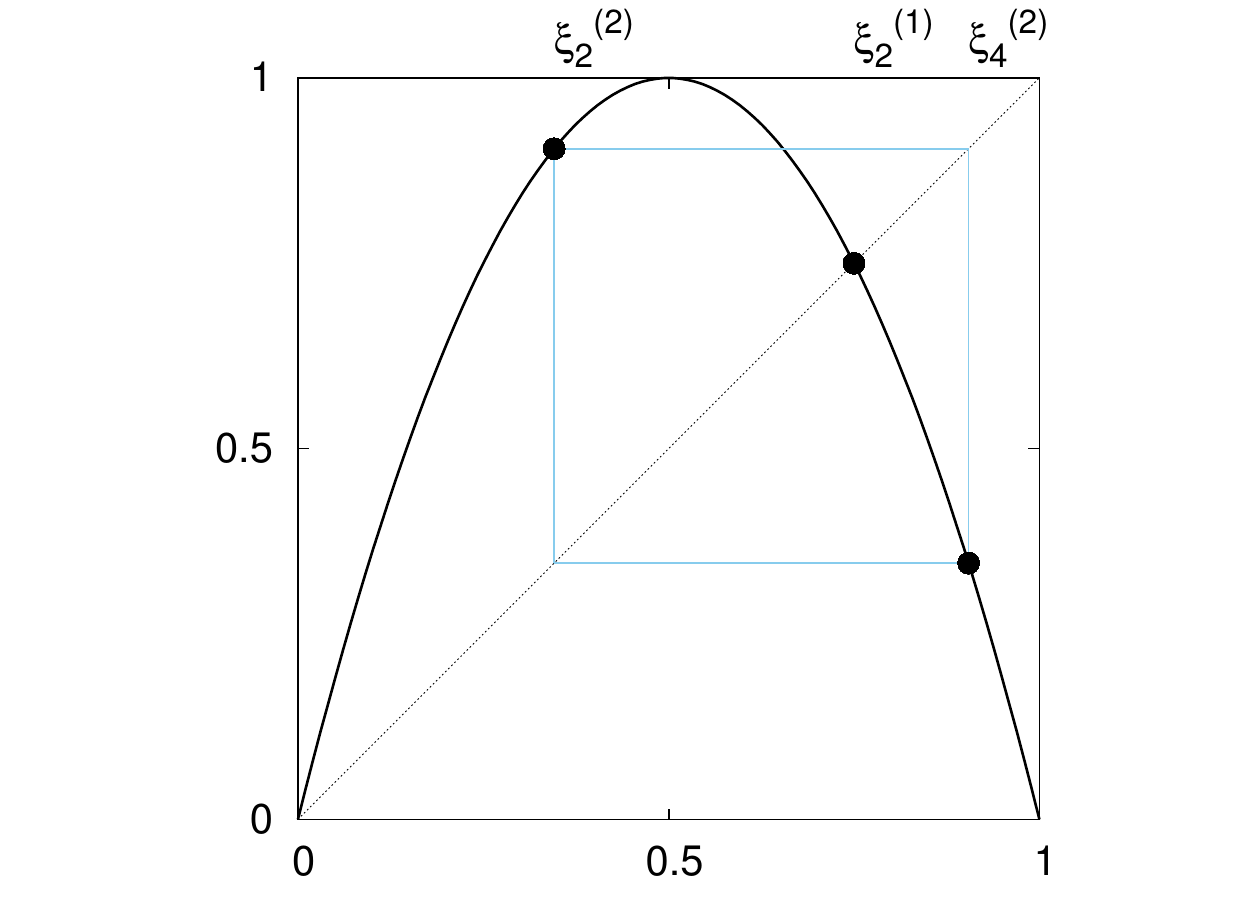}
\caption{Fully chaotic logistic map with nontrivial fixed point
$\xi_2^{(1)}$ and unstable period two orbit $\xi_2^{(2)}$, $\xi_4^{(2)}$.}
\label{per2}
\end{figure}

By construction, a motif of length two requires $x_0>x_1=f(x_0)$ and
$x_0<x_2=f(x_1)$. Clearly $x_0\in [\xi_2^{(1)},1]$. Furthermore
$x_0$ cannot exceed the largest of the period two points,
$\xi_4^{(2)}$ as otherwise the image of $x_1$ would be smaller
than $\xi_4^{(2)}$, and hence smaller than $x_0$. In addition, values $x_0$
smaller than $\xi_4^{(2)}$ result in images $x_1=f(x_0)$ which on a further
iteration step give a value exceeding $x_0$, see as well figure \ref{per2}.
Thus the relevant initial conditions for the motif form again a single interval 
$${\cal I}_0^{(2)}=[\xi_3^{(2)},\xi_4^{(2)}]$$
bounded by the two largest fixed points of the second iterate, if we keep
in mind that $\xi_3^{(2)}=\xi_2^{(1)}$.

\noindent There are no further motifs (i.e. motifs with hidden nodes)
which result in an \textit{out}-degree $k^{\text{out}}(x)=2$, and the reason is simple: if we had a hidden node in such a structure, it would require
iterates of the form $x_0>x_1>x_2$. But, again, the logistic map does not allow for
a time series of three consecutive decreasing values. $x_0>x_1$ requires
$x_0\in [\xi_2^{(1)},1]$ so that $x_1\in [0,\xi_2^{(1)}]$. If $x_1<1/2$ then
the image obeys $x_2>x_1$ as the logistic map is increasing. If $x_1\geq 1/2$
then $x_2 \in [\xi_2^{(1)},1]$ and again $x_2>x_1$.\\

\noindent We conclude that the single motif with $p=2$ captures at once all cases with out-degree two and therefore $k^{\text{out}}(x)=2$ on the 
interval ${\cal I}_0^{(2)}$ (see figures \ref{part_k} and \ref{zoom}).

\subsubsection{$p=3$}
\begin{figure}[h]
\centering
\includegraphics[width=0.99\columnwidth]{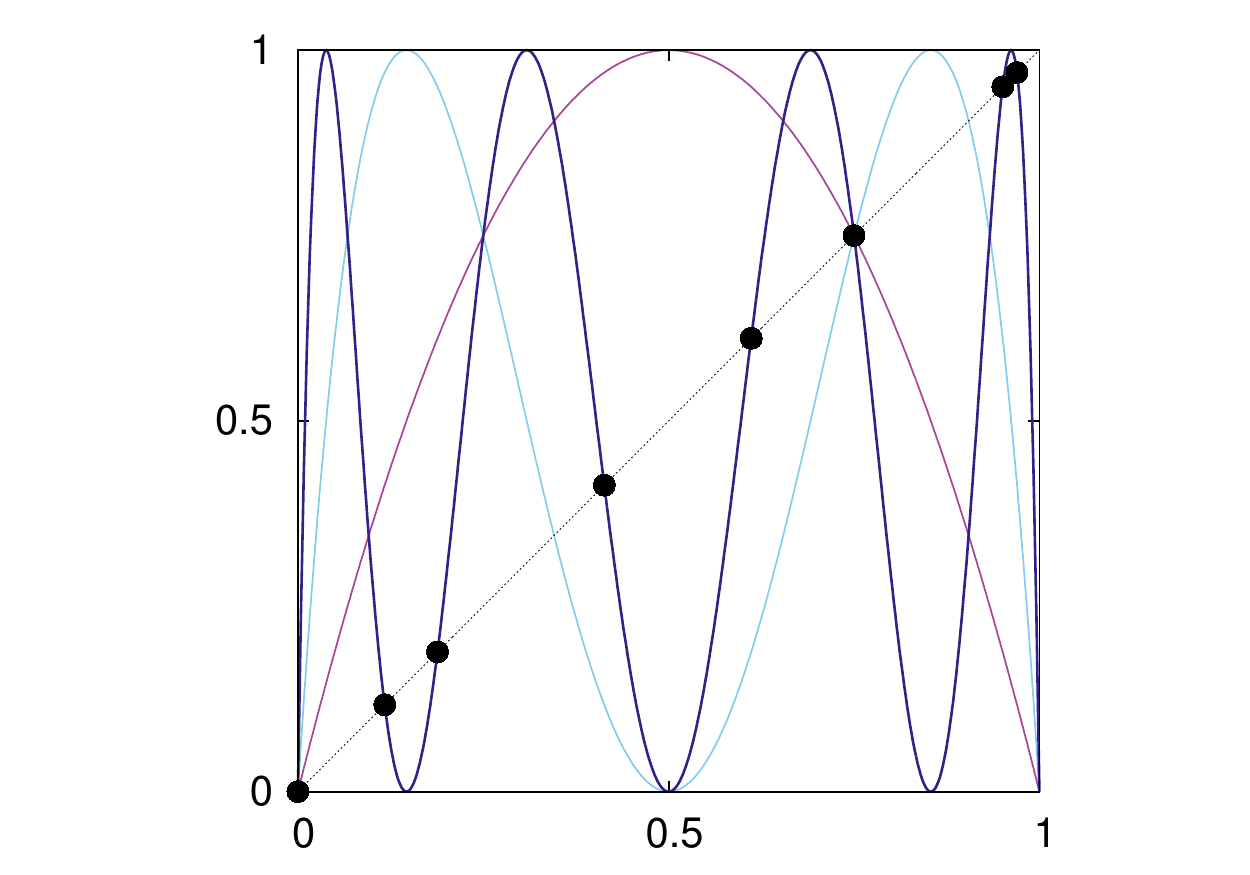}
\caption{Third iterate (blue, thick line) and first and second iterate (red and cyan, thin lines)
of the fully chaotic logistic map. Full symbols indicate the 8 period-three
points, $\xi_k^{(3)}$.}
\label{per3}
\end{figure}

Since a decreasing sequence of three consecutive nodes
is forbidden, the initial part of the motif has to obey
$x_1<x_2<x_0$. With the additional and necessary condition $x_0\leq x_3$, these
inequalities determine the motif uniquely, which therefore lacks hidden
nodes (see figure \ref{motive}). To compute the set of initial conditions
that gives rise to this motif, observe the necessary condition $x_0\leq x_3=f^{(3)}(x_0)$. Hence, the set
has to be contained in those intervals where the third iterate exceeds the
diagonal, see figure \ref{per3}. These four intervals are bounded by
periodic points of order three, $[\xi_{2k-1}^{(3)},\xi_{2k}^{(3)}]$
where $1\leq k\leq 4$.  Within the rightmost interval
bounded by the largest and the second largest periodic point,  the 
lower order iterates $f^{(m)}(x)$ with $m=1,2$
have a well defined order, i.e. their graphs do not cross (for a formal
proof in the general setting see below), and the
iterates follow the order specified by the motif, 
$f(x)<f^{(2)}(x)<x<f^{(3)}(x)$. For all the other intervals the third iterate
is not even visible as a lower iterate exceeds the initial value, 
$f(x)>x$.
Hence, the set of initial conditions giving rise to the unique motif of length
$p=3$ is given by $${\cal I}_L^{(3)}=[\xi_7^{(3)},\xi_8^{(3)}].$$ While this set and the
corresponding motif gives rise to a node with \textit{out}-degree three, there are
now additional motifs and initial conditions which will result 
in the same node degree, as we will discover shortly.

\subsubsection{$p=4$}
\begin{figure}[h]
\centering
\includegraphics[width=0.99\columnwidth]{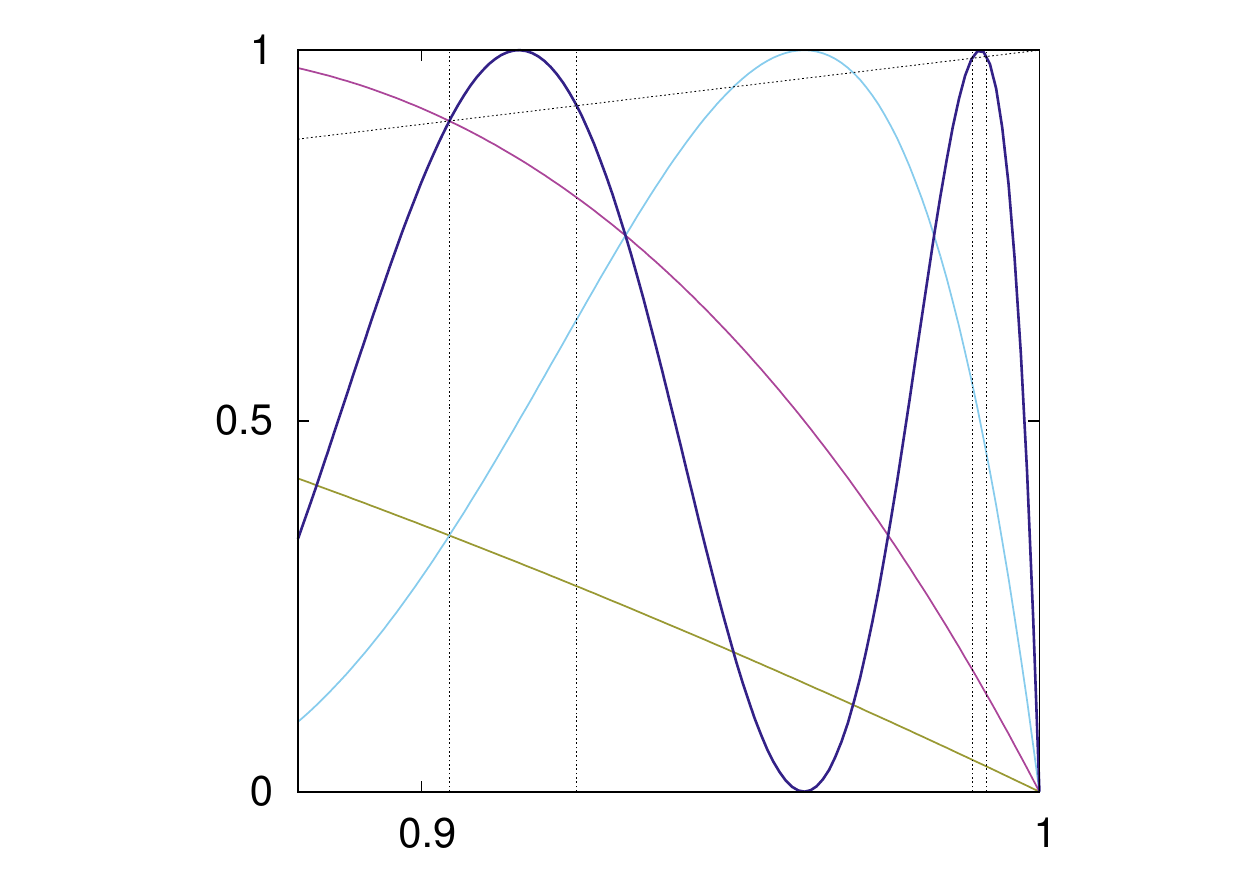}
\caption{Fourth iterate (blue, thick line) and lower order iterates (bronze, red and cyan, thin lines)
of the fully chaotic logistic map, in the region beyond the period two
orbit. Vertical lines indicate intervals determined by the largest
period four points $\xi_k^{(4)}$.}
\label{per4}
\end{figure}

Since a necessary condition for a motif of length $p=4$ is given by
$x_0\leq f^{(4)}(x_0)$ we will look at the graph of the iterates $f^{(k)}$ for
$k=1,2,3,4$, see figure \ref{per4}. As in the previous case the condition
$x\leq f^{(4)}(x)$ determines eight intervals bounded by points of
period four, $[\xi_{2k-1}^{(4)},\xi_{2k}^{(4)}]$ with $1 \leq k \leq 8$. 
Within any interval which obeys the necessary condition
$x\leq f^{(4)}(x)$ the lower order iterates $f^{(m)}(x)$, $m=1,2,3$
have a well defined order, i.e., their graphs do not cross. Otherwise we 
would have $f^{(m_1)}(x)=f^{(m_2)}(x)$ for some value $x$ in the interval,
i.e., a periodic orbit of period $|m_2-m_1|<4$. Then however $x_4$
cannot not be visible, as its value would be taken by 
one of the previous iterates.\\

\noindent It is obvious that we only need to consider the region 
beyond the period-two orbit, $x>\xi_2^{(2)}$, as otherwise
either $f(x)>x$ or $f^{(2)}(x)>x$.
Only the two largest intervals $${\cal I}_{LL}^{(4)}=[\xi_7^{(4)},\xi_8^{(4)}]$$
and $${\cal I}_{LR}^{(4)}=[\xi_5^{(4)},\xi_6^{(4)}]$$
obey the visibility constraints for intermediate nodes, i.e., 
$f^{(k)}(x_0)=x_k<x_0$ for $1\leq k \leq 3$ (see the section below
for the notation used to label the intervals).
These intervals give rise to
two motifs, see figure \ref{motive}, with the nodes ordered according to
$x_1<x_2<x_3<x_0<x_4$ or $x_1<x_3<x_2<x_0<x_4$. It is in fact rather
obvious that only two motifs exist as the three initial nodes have to
obey $x_1<x_2<x_0$ as already mentioned above. Thus, there are just two
possibilities left for the visibility of the third node. The case $p=4$ is 
the first instance of a motif with a hidden node: one motif yields   
\textit{out}-degree three, while the other one yields out-degree four.\\

\noindent In fact, there are no further motifs resulting in an \textit{out}-degree three. Hence,
the set of initial conditions giving rise to \textit{out}-degree three
consists of two disjoint intervals $$\{x_0| k^{\text{out}}(x_0)=3\}={\cal I}_L^{(3)}\cup {\cal I}_{LR}^{(4)},$$ see as well
figure \ref{zoom}. To prove this claim we need to show that for a sequence
of nodes with $x_1<x_3<x_2<x_0$ the following point $x_4$ cannot be
a hidden node with $x_4<x_2$. Since $x_3<x_2$ we know, see the discussion
of the logistic map in the previous section, that $x_2$ is contained in 
$[\xi_2^{(1)},1]$, i.e.,
$x_2$ is bounded from below by the nontrivial fixed point $x_2>\xi_2^{(1)}$. 
Since
$x_2<x_0$ the graph of the second iterate $f^{(2)}$ (see, 
e.g., figure \ref{per3})
tells us that $x_0>\xi_4^{(2)}$ and hence $x_2=f^{(2)}(x_0)<\xi_4^{(2)}$.
Thus $x_2$ is contained in the interval $[\xi_2^{(1)},\xi_4^{(2)}]$ but
on this interval the graph of the second iterate is above the diagonal,
$f^{(2)}(x)>x$. Hence $x_4>x_2$.\\

\noindent motifs of length $p>4$ become increasingly difficult
to construct by explicit methods. Thus, a more systematic approach is needed.

\subsection{motifs without hidden nodes}

Before we address the general case, let us first focus on motifs of
length $p$ where all nodes are visible (no hidden nodes), i.e. $x_1<x_2<\ldots<x_{p-1}<x_0<x_p$. In other words, in these motifs all intermediate nodes constitute an increasing subsequence. In what follows we analytically obtain the sets of initial conditions yielding these motifs.\\ 

\noindent We start by observing that the $p$-th iterate $f^{(p)}$ is a $p$-modal function with $2^{p-1}$ maxima 
at one and
$2^{p-1}+1$ minima at zero. Between extrema, branches are monotonic.
The $p$-th iterate has $2^p$ fixed points, $\xi^{(p)}_k$ where
$1\leq k\leq 2^p$, and given the properties mentioned before the
inequality $f^{(p)}(x)>x$ is satisfied on the $2^{p-1}$ intervals
$[\xi_{2k-1}^{(p)}, \xi_{2k}^{(p)}]$ where $1\leq k \leq 2^{p-1}$. As already
shown above, on any interval which obeys the visibility condition
$x_0\leq x_p=f^{(p)}(x_0)$ the lower order iterates
$f^{(m)}$ with $1\leq m \leq p-1$ have a well defined order, i.e.,
these functions do not intersect. In what follows we will only
consider these intervals.\\

\noindent The ordering of the branches of the iterates is different for
different intervals $[\xi_{2k-1}^{(p)}, \xi_{2k}^{(p)}]$. To prove this claim
we need to resort to symbolic dynamics.
Within $[\xi_{2k-1}^{(p)}, \xi_{2k}^{(p)}]$ there is a single point
$x_m$ where $f^{(p)}$ takes its maximum, i.e., $f^{(p)}(x_m)=1$. The orbit
of this point, i.e. the sequence $f^{(k)}(x_m)$ with $1\leq k\leq p-1$
determines the ordering of the branches. Similarly, if the ordering 
of the branches is given we can compute the
value of $x_m$ by backward iteration.
The condition $f^{(p)}(x_m)=1$ means  $f^{(p-1)}(x_m)=1/2$. The value of
$f^{(p-2)}(x_m)$ is the preimage of $1/2$. The relative ordering of
the iterates $f^{(p-2)}$ and $f^{(p-1)}$ tells us whether
$f^{(p-2)}(x_m)$ is smaller or larger than  $f^{(p-1)}(x_m)=1/2$, i.e.,
it tells us whether to apply the left of the right branch of the inverse 
function $f^{-1}$. Hence we can uniquely compute the entire sequence 
$f^{(k)}(x_m)$ as the relative ordering of the iterates $f^{(k)}$ and 
$f^{(p-1)}$ tells us in each step
which branch of the inverse function has to be applied. That in turn implies 
that two different intervals must have two different orderings of iterates.
Otherwise one would obtain the same value for $x_m$, but the two intervals
do not have a point in common. In summary, we can label each of the intervals
$[\xi_{2k-1}^{(p)}, \xi_{2k}^{(p)}]$ by a symbol sequence of L's and R's such 
that $f^{(k)}(x_m) \in {\cal I}_{L/R}$ for $1\leq k\leq p-2$. This notation
has been already used in the previous discussion of special cases (see
figure \ref{motive}).\\

\noindent With these preliminary considerations we are now able to
evaluate motifs without hidden nodes $x_1<x_2<\ldots<x_{p-1}<x_0<x_p$.
In this case the corresponding orbit $x_m,f(x_m),\ldots,f^{(p-2)}(x_m),
f^{(p-1)}(x_m)=1/2,f^{(p)}=1$ has a monotonic increasing part
$f(x_m)<f^{(2)}(x_m)<\ldots<f^{(p-2}(x_m)<1/2$, meaning that all the orbit
points are preimages of $1/2$ by using the left branch of the inverse
function. Hence, the value obtained for $f(x_m)$ is the smallest possible value
among all the intervals $[\xi_{2k-1}^{(p)}, \xi_{2k}^{(p)}]$ and in turn
$x_m$ is the largest possible value. Thus the interval 
giving the motive without hidden nodes is the rightmost
interval  $[\xi_{2^p-1}^{(p)}, \xi_{2^p}^{(p)}]$. According to our notation
it is labelled by ${\cal I}^{(p )}_{LL\ldots L}$ (see as well figure \ref{motive}). Finally we have shown that
$${\cal I}^{(p )}_{LL\ldots L}=[\xi_{2^p-1}^{(p)}, \xi_{2^p}^{(p )}].$$
As these motifs lack hidden nodes, all initial conditions in ${\cal I}^{(p )}_{LL\ldots L}$ have an associated node with \textit{out}-degree $p$.\\

\paragraph*{The weight of motifs without hidden nodes:}  To figure out which part of the phase space is covered by motifs without hidden nodes let us consider in more in detail the intervals ${\cal I}^{(p )}_{LL\ldots L}$. Denoting the length of an interval by $| [a,b] |=|b-a|$ the ratio 
\begin{equation}
\zeta(p)=\frac{|{\cal I}^{(p )}_{LL\ldots L}|}{|{\cal I}^{(p-1)}_{LL \ldots L}|}, 
\ (p\geq 2)
\end{equation}
measures the shrinking rate of the intervals 
of initial conditions resulting in motifs of length $p$
without hidden nodes. Interestingly, numerical evidence suggests that this rate rapidly converges 
to  
\begin{equation}
\lim_{p\to \infty}\zeta(p)\approx  0.125 \, . \nonumber
\end{equation}
One can compute this limit analytically building on the conjugation to the tent map, which tells us that
$$\xi_{2^p}^{(p )}=\sin^2\bigg(\frac{\pi}{2}\frac{2^p}{2^p+1}\bigg);\ \xi_{2^p-1}^{(p )}=\sin^2\bigg(\frac{\pi}{2}\frac{2^p-2}{2^p-1}\bigg) \, .$$
Therefore
$$\lim_{p\to \infty}\zeta(p) = \lim_{p\to \infty}\frac{\sin^2\bigg(\frac{\pi}{2}\frac{2^p}{2^p+1}\bigg) - \sin^2\bigg(\frac{\pi}{2}\frac{2^p-2}{2^p-1}\bigg)}{\sin^2\bigg(\frac{\pi}{2}\frac{2^{p-1}}{2^{p-1}+1}\bigg)-\sin^2\bigg(\frac{\pi}{2}\frac{2^{p-1}-2}{2^{p-1}-1}\bigg)}=\frac{1}{8} \, .$$
Similarly, if $\mathbb{P}^{(0)}(p)$ defines the probability of 
finding a certain symbol $k^{\text{out}}=p$ without hidden nodes, then
\begin{eqnarray}
&&\mathbb{P}_0 (p )=\int_{\xi_{2^p-1}^{(p )}}^{\xi_{2^p}^{(p )}}\frac{dx}{\pi \sqrt{x(1-x)}}=\nonumber \\
&&\frac{ \arcsin(2 \xi^{(p)}_{2^p} -1 ) - 
\arcsin(2\xi^{(p)}_{2^p-1}-1) }{\pi} = \frac{2}{4^p-1} \nonumber
\end{eqnarray}
giving a rigorous lower bound on the exponential decay of the degree
distribution of the HVG.
Accordingly, the total contribution of motifs without hidden nodes, 
$\Lambda_0$, can be defined as
\begin{equation}
\Lambda_0= \sum_{p=1}^\infty \mathbb{P}_0 (p )=\sum_{p=1}^\infty  
\frac{2}{4^p-1} \, 
\label{qpoly}
\end{equation}
This series can be written in terms of the
q-polygamma function, and the expression has
some similarity to the Erd\"os-Borwein constant \footnote{T. Prellberg, private communication.}. The value of $\Lambda_0$ is irrational by a theorem of Borwein \cite{borwein}, but apart from this fact not much seems to be known
about this constant. We find that $\Lambda_0\approx 0.842195\dots$, converging after $p=11$.  
This means that up to $84\%$ of all initial conditions generate trajectories whose associated \textit{out}-degree sequence lacks hidden nodes.

\subsection{General motifs, periodic points, and symbolic encoding}

To give a complete description of motifs of length $p>2$ we need a few more details
about symbolic dynamics. To keep the presentation self-contained
we first summarise a few facts for the convenience 
of the reader, even though more comprehensive reviews can
be found in textbooks \cite{CE}.\\

\noindent Given the canonical partition of the interval in terms of two subintervals ${\cal I}_L=[0,1/2]$ and
${\cal I}_R=[1/2,1]$ we can (essentially) associate to each orbit of the map $(x_0,x_1,\ldots)$,
that means to each initial condition $x_0$,  a symbol sequence 
$\sigma_0,\sigma_1,\ldots$ consisting of symbols $\sigma_k\in\{L,R\}$,
such that the symbols tell us the location of the orbit points,
$x_k\in {\cal I}_{\sigma_k}$. For the case of the
fully chaotic logistic map all symbol sequences are indeed admissible. Other 
parameter values can be covered as well by pruning the set of admissible 
symbol sequences.\\

\paragraph*{Ordering of symbol sequences:} Given two initial conditions $x_0$ and $\bar{x}_0$ for which $x_0<\bar{x}_0$, this usual {\it order} in phase space $x_0<\bar{x}_0$ induces a corresponding order 
$\prec$ among symbol sequences, which is essentially a binary order taking into
account that one branch of the logistic map is decreasing. To be slightly
more specific, such ordering relation is defined as follows:\\
(i) $L \sigma_1 \sigma_2 \ldots \prec R \bar{\sigma}_1
\bar{\sigma}_2 \ldots$. Furthermore, \\ (ii) if the first $N$ symbols of
two sequences coincide $\sigma_0 \ldots \sigma_{N-1}=\bar{\sigma}_0 
\ldots \bar{\sigma}_{N-1}$ and if this string contains an even number of R's,
then $\sigma_0 \ldots \sigma_{N-1} L \ldots \prec 
\sigma_0 \ldots \sigma_{N-1} R \ldots$.
Otherwise, \\(iii) if the string contains
an odd number of R's then $\sigma_0 \ldots \sigma_{N-1} R \ldots \prec 
\sigma_0 \ldots \sigma_{N-1} L \ldots$.\\

\paragraph*{Maximal sequences:} This ordering of symbol sequences is defined in such a way that it coincides with the order of the corresponding 
initial conditions. Periodic symbol sequences, $\sigma_{k+p}=\sigma_k$ for
$k\geq 0$, correspond to phase space points of period $p$. 
We will use the notation $\overline{\sigma_0\ldots \sigma_{p-1}}$ to
denote periodic symbol sequences. A 
periodic sequence $\overline{\sigma_0\ldots \sigma_{p-1}}$
is said to be a {\it maximal symbol sequence} if
$\sigma_{k} \sigma_{k+1}\ldots \prec \sigma_0 \sigma_1 \ldots$ for $1\leq k
\leq p-1$. In geometric terms a maximal periodic symbol sequence corresponds
to a period $p$ point $x_0$ of the map, such that $x_0$ is the largest value
among all iterates, $x_0>x_k$ for $1 \leq k \leq p-1$.\\ 

\paragraph*{Encoding of motifs with length $p>2$:} Using symbolic dynamics we are now equipped with the necessary tools to rephrase the results which have been
implicitly obtained in the previous sections. For a given motif
of length $p$ with a 
given order of nodes the corresponding initial conditions $x_0$ are contained
in an interval $[\xi_{2k-1}^{(p)}, \xi_{2k}^{(p)}]$ whose endpoints are
periodic points of order $p$. If $x_0 \in [\xi_{2k-1}^{(p)}, \xi_{2k}^{(p)}]$
scans the interval the iterates $x_k=f^k(x_0)$ for $1\leq k \leq p-2$ never 
become $1/2$, i.e., we can attach a unique symbol $\sigma_k \in \{L,R\}$ 
to $x_k$ according to $x_k\in {\cal I}_{\sigma_k}$. When $x_0$ scans the interval 
the iterate $x_{p-1}=f^{(p-1)}(x_0)$ crosses $1/2$ once, meaning that the 
corresponding symbol $\sigma_{p-1}$ changes. Finally the visibility 
conditions ensure that
$x_0$ and $x_p$ are contained in ${\cal I}_R$ (recall that we consider the case
$p>2$) so that $\sigma_0=\sigma_p=R$. Hence the two periodic points which
are the boundaries of the interval of the motif have symbol code
$\overline{R\sigma_1\ldots \sigma_{p-2} L}$ and 
$\overline{R\sigma_1\ldots \sigma_{p-2} R}$ where the finite symbol string
$\sigma_1\ldots \sigma_{p-2}$ is the unique identifier of the motive
\begin{equation}\label{Imtv}
{\cal I}^{(p )}_{\sigma_1\ldots \sigma_{p-2}}=[\xi_{2k-1}^{(p)}, \xi_{2k}^{(p)}], \quad
(p>2) \, .
\end{equation}
The two periodic symbol strings are maximal sequences, since visibility
requires that $x_0=x_p$ is the largest value in the respective periodic
orbit.\\ 

\paragraph*{Properties of $k^{\text{out}}(x)$:} The theoretical analysis displayed above give us a workable solution to label all motifs in terms of subintervals, i.e. we are able to associate to each motif a set of initial conditions. That is, we have been able to make a partition of the phase space into a countable union of subintervals, and we indeed control the location of each of the (countably infinite) subintervals. Moreover, $k^{\text{out}}(x)$ takes a constant value on each of these subintervals. However, there does not seem to be a simple recipe relating
the properties of the symbol string with the actual \textit{out}-degree of the specific
motif. In particular, that means that we can find many different (disjoint) subintervals whose initial conditions yield the same \textit{out}-degree. Because of this, the full explicit construction of $k^{\text{out}}(x)$ is currently out of reach. 
Notwithstanding, we are able to explore some further properties of this 
function, as follows.\\

\begin{figure}[h]
\centering
\includegraphics[width=0.99\columnwidth]{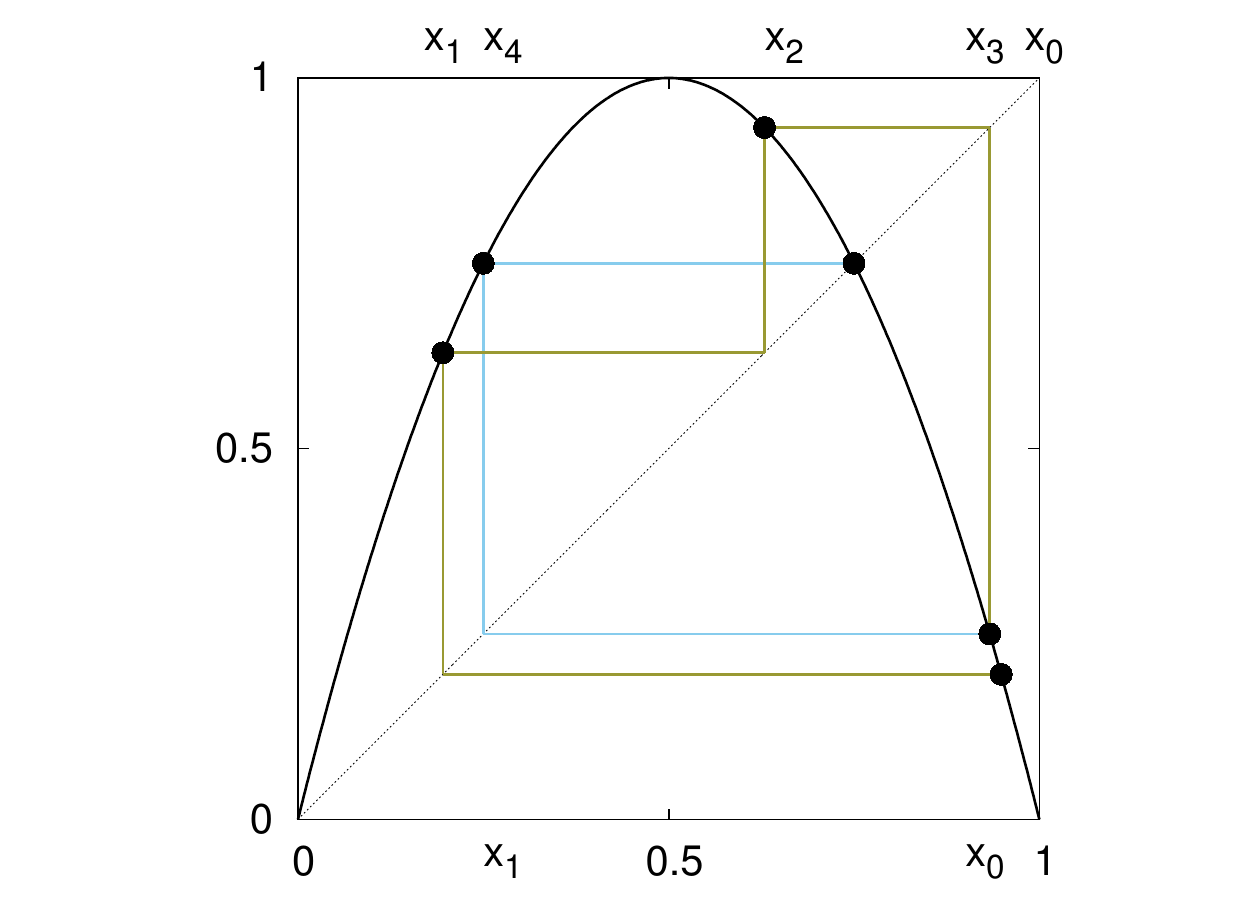}
\caption{Orbits related to motifs with an arbitrary large number of
nodes (cyan, bottom labels) and motifs with
an arbitrary large number of hidden nodes (cyan and bronze, top labels).}
\label{orb}
\end{figure}

\noindent We first claim that the node degree of motifs is not bounded and the function $k^{\text{out}}(x)$
can take arbitrarily large values in
any small neighbourhood of particular $x$ values. The simplest case
has been considered in the previous section but there are less
trivial cases. Consider the orbit $x_0,x_1,\xi_2^{(1)},
\xi_2^{(1)},\ldots$ which ends up in the nontrivial fixed point after
two iteration steps (see figure \ref{orb}), 
i.e., $x_0=1/2+\sqrt{3}/4=.9330\ldots$. 
Changing the value of $x_0$ by a very small amount the orbit will slowly be 
repelled from
the unstable fixed point so that nodes $x_0$, $x_1$ and every other of
the following iterates will be visible until finally the motive terminates 
with a value exceeding $x_0$. By making the increment as small as we wish 
we can  make the node degree as large as we want. Hence $k^{\text{out}}(x)$ 
is unbounded at $x=1/2+\sqrt{3}/4$ as can be seen as well in figure \ref{zoom}. 
We can easily construct a countable infinite set of such $x$-values, related 
to unstable orbits of higher period.\\ 

\noindent Second, we then claim that there are motifs with an arbitrarily large number of hidden nodes.
Consider for instance the orbit of $x_0=0.9484\ldots$ which ends up in the 
nontrivial 
fixed point $\xi_2^{(1)}=3/4$ after five iterations and whose transient obeys the pattern
$x_1<x_4<x_2<x_3$ see figure \ref{orb}. Nodes $x_1$, $x_2$ and $x_3$ 
are visible (cf. ${\cal I}_{LL}^{(4)}$ in figure \ref{motive}) and $x_4$ and the 
fixed point are invisible. If we change $x_0$ by a small amount the 
orbit will spiral around
the unstable fixed point generating a large number of invisible nodes
in the motive, until finally the motive terminates with a value larger
than the initial value. As in the previous case in any open neighbourhood of
 $x_0=0.9484\ldots$ we can find motifs with as many hidden nodes as we want.
In particular, our argument implies that
there is a countably infinite set of motifs with \textit{out}-degree $p\ge 4$.
Hence, the set of values where $k^{\text{out}}(x)=4$ consists of a 
countably infinite union
of intervals, see figure \ref{zoom}, and there is no obvious way to
characterise this set.

\section{Discussion}

In this work we have explored the properties of the degree sequence of 
horizontal visibility graphs associated with chaotic time series using 
symbolic dynamics. For concreteness, we have focused on the 
logistic map $x_{t+1}=r x_t (1-x_t)$, a canonical interval map showing 
transition from regular to chaotic dynamics as $r$ is varied. Numerically, we have shown that the Lyapunov exponent of the logistic map 
is well approximated asymptotically by a sequence of block entropies on both 
the degree and \textit{out}-degree sequences. Via Pesin theorem, this suggests that this sequence of entropies is 
converging to the Kolmogorov-Sinai entropy of the map, and therefore 
constitutes a combinatoric version of the metric dynamical invariant. 
Furthermore, this connection suggests that the horizontal visibility graph 
is inducing a symbolic dynamics and effectively produces partitions of the 
interval which could be generating. Note that the algorithm itself does not pre-define the alphabet (i.e. the number of different degrees), and the way it constructs the degree sequence from the original time series does not suggest a priori that there might be an underlying partition of the phase space operating at all. So an explicit construction of such partition would constitute an unexpected and nontrivial result.\\

\begin{figure}[h]
\centering
\includegraphics[width=0.95\columnwidth]{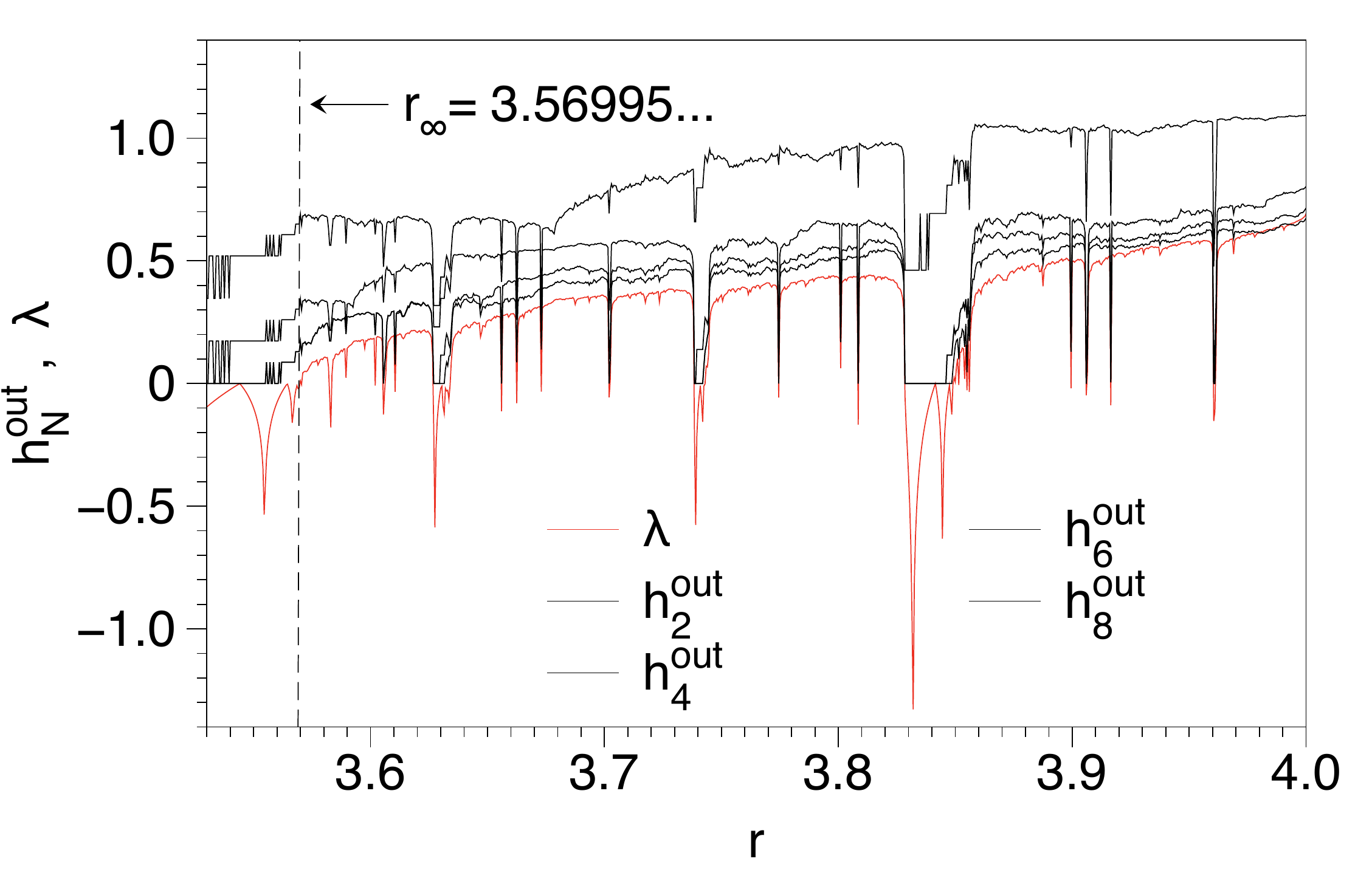}
\caption{$h^{\text{out}}_N$ for different block sizes $N$, as a function 
of the map's parameter $r$, where we appreciate how the successive 
approximations to the entropy rate converge to the Lyapunov 
exponent $\lambda$ of the map, for all values of $r$.}
\label{BEout}
\end{figure}

\noindent To further explore this possibility, we have elaborated on the explicit construction of such effective partition in a sequential way.
Unlike the degree, we have shown that the \textit{out}-degree
is a well defined phase space function and the level curves, i.e., the
sets on which the \textit{out}-degree takes a specific value, provide a {\it countable
partition} of the phase space. Level sets for \textit{out}-degree one, two, and three
are either intervals or the union of two intervals, while all sets of
degree larger than three consist of a countable infinite union of subintervals. We have proved that all subintervals
are bounded by periodic points and each subinterval can be labelled according
to a so-called maximal symbol sequence. We found that the vast majority of the phase space is
covered by sets with corresponding low \textit{out}-degree or by motifs without
any hidden nodes, where explicit calculations are possible.\\

\noindent Furthermore, the entropy based on the out-degree sequence, eq.(\ref{entout}), shows
a striking similarity to the Kolmogorov-Sinai entropy (see figure \ref{BEout}). 
In fact, the \textit{out}-degree
provides a partition of the phase space, and considering out-degree
sequences implicitly provides a dynamic refinement of this partition,
as in the case of the Kolmogorov-Sinai entropy. The set of phase space points
giving rise to a finite degree sequence $k^{\text{out}}_1,k^{\text{out}}_2,
\dots, k^{\text{out}}_N$ is the intersection of the sets such that
$f^{(m-1)}(x)$ is contained in the part with \textit{out}-degree $k^{\text{out}}_m$ 
for $1\leq m\leq N$. Such a construction
is precisely the definition of a dynamically refined partition.
A formal proof along standard lines, to show that the entropy based
on degree distributions equals the Kolmogorov-Sinai entropy, would amount
to establish the generating property of the underlying partition.
However the partition defined by the out-degree can hardly be generating in the
topological sense as the map is non monotonic on the part ${\cal I}^{(1)}_0$.
Nevertheless, \textit{out}-degree sequences can be efficiently used to count periodic
orbits and thus share properties of a generating partition. Obviously
any periodic orbit results in a periodic degree sequence of the same period.
In addition, we have shown that any motif ${\cal I}^{(p)}_{\underline{\sigma}}$ has a
characteristic ordering of branches of iterates, and this ordering is 
different for each motif. The ordering of these branches determines the
degree sequence, so that the degree sequence is a fingerprint of the motif.
If we confine to periodic degree sequences and periodic orbits we have thus 
shown that a periodic degree sequence is a specific property of the two periodic
orbits which constitute the boundary points of the motif, i.e., the mapping from degree sequences to periodic orbits is one to two.\\

\noindent Hence, the entropy based on \textit{out}-degrees is closely related, but not identical,
to well established concepts in dynamical systems theory. A similar
statement is of course valid for the undirected degree sequence or for 
the \textit{in}-degree sequences. From a rigorous perspective their 
relation to topological properties
of the underlying dynamics remains as an open problem, even though we have compelling numerical and analytic evidence that all the entropy values coincide. 
Other interesting open questions include the extension of this 
technique to chaotic maps in higher dimensions \cite{Enzo}.\\

\begin{acknowledgments}
\vspace{-2mm}
We thank Thomas Prellberg for pointing out the simiarity of $\Lambda_0$ to the Erd\"os-Borwein constant. LL acknowledges funding from EPSRC Early Career Fellowship EP/P01660X/1.
\end{acknowledgments}

\bibliography{apssamp}

\end{document}